\documentclass[twocolumn,aps,prl,groupedaddress]{revtex4}
\usepackage{graphicx}
\usepackage{dcolumn}
\usepackage{bm}
\usepackage{braket}
\begin{document}

\title{Spin-momentum locking and Majorana fermions in charge carrier hole epitaxial wires}
\author{G.E. Simion}
\thanks{Now at IMEC , Kapeldreef 75, 3001, Leuven, Belgium}
\author{Y.B. Lyanda-Geller}
\email{yuli@purdue.edu} \affiliation{Department of Physics and Astronomy and Purdue Quantum Institute, Purdue University, West Lafayette IN, 47907 USA}

\date{April, 10 2019}

\begin{abstract}
Epitaxial semiconductor nanowires with charge carrier holes can exhibit an  infinite mass of holes and spin-locking due to chiral spectrum linear in momentum and spin. The criterion for emergence of topological superconductivity and Majorana fermions in these wires coupled to an s-type superconductors is the same as in topological insulators, and opposite to the criterion of onset of Majorana modes in quantum wires with parabolic spectrum in the presence of spin-orbit interactions.
\end{abstract}
\maketitle

Quantum wires proximity-coupled to a superconductor are one of the important settings for Majorana fermions, particles with
non-Abelian statistics paving the way to topological fault-tolerant quantum computing.
Of particular interest are wires with charge carrier holes, which promise to have strong spin-orbit interactions.
 These wires have various interesting properties, and are thought for applications in spintronics and quantum information science.

There are two types of wires with charge carrier holes. Wires based on quantum wells, which are obtained from wells lithographically or via electrostatic gating, as well as cleaved edge overgrowth wires belong to the first type, in which quantization
along one of the directions perpendicular to the wire is much stronger than quantization in the other direction. Epitaxial or core-shell nanowires comprise
a second type, in which size quantization in two directions perpendicular to the direction of free propagation of holes is comparable. Epitaxial and core shell nanowires attracted considerable attention recently. Epitaxial wires can form heterostructures with superconductor such as Al leading to superconducting proximity effect.

Theoretical treatment of low-dimensional holes have been controversial for a long time. It is important to recognize that the effect of mutual transformation of heavy and light holes upon reflection from the heteroboundaries of the quantum well generally cannot be taken into account perturbatively. This understanding came with the work of  Nedorezov  \cite{nedorezov1971space}, but was seldom applied afterwards \cite{merkulov1991momentum,rashba1988spin}.
It was almost ignored over the past two decades, when holes were largely treated as electrons,\cite{Arovas,LyandaGeller2004,LossPRL05,Bernevig,Hughes,mao2012hole}. The nonperturbative approach, however, is important for determination of the effective masses, g-factors and spin-orbit constants, as discussed recently
\cite{simion2014magnetic,Ivchenko,Jingcheng}. In particular, mutual transformation of heavy and light holes is important in core-shell and epitaxial nanowires \cite{Losswires,Klinovaja,Zulicke}.

In the present paper we show that hole spectrum in epitaxial nanowires can be characterized by a wide range of values of the effective hole mass, including an  infinite value. In this case there is locking of spin to momentum, because the principal term in Hamiltonian of the wires becomes chiral, linear in momentum and spin.
As a result, the criterion for the emergence of Majorana fermions and topological superconductivity becomes the same as in topological insulators in proximity of a superconductor \cite{fu2008superconducting,alicea2012new}, and topological superconductivity and Majorana fermions in these wires can emerge in small and even zero magnetic
field. This is in contrast to large magnetic fields with Zeeman energy comparable to superconducting gap and chemical potential in spin-orbit quantum wires with parabolic spectrum \cite{lutchyn2010majorana,oreg2010helical}. We illustrate the emergence of an infinite mass in a model case
with hard-wall boundary conditions. The obtained phase diagram implies that for certain solid solutions or in the presence of strain the infinite hole mass in wires can show up experimentally.

The Luttinger Hamiltonian for holes is
\begin{eqnarray}
\label{eq:Luttinger} \hat
H_L\!&\!=\!&\!\frac{\hbar^2}{2m_0}\!\left[\!\left(\gamma_1+\frac{5 \gamma_2}{2}\right)
\hat{\bf{k}}^2I\!-\!2\gamma_2(\hat k_x^2J_x^2+\hat k_y^2J_y^2+\hat k_z^2J_z^2)\right.\nonumber\\
\!&\!-\!&\left.\!4\gamma_3\!\left(\hat k_x \hat k_y\{J_x,J_y\}\! +\! \hat k_y \hat
k_z\{J_y,J_z\}\!+\!\hat k_z\hat k_x\{J_z,J_x\}\right)\right]\nonumber\\.
\end{eqnarray}
where $J_x, J_y,J_z$ are the $3/2$ angular momentum matrices, and $\gamma_1$,
$\gamma_3$, $\gamma_3$ are Luttinger parameters. Holes are confined
to a cylindrical wire of radius $R$ and  $z$-direction is the wire axis.
The hole wavefunctions satisfy the boundary condition $\Psi(R,\phi,z)=0$.
%The wavevector along the $z-$ direction $k_z$ is a good quantum number.

We employ the axial approximation, in which the terms containing
$(\gamma_3-\gamma_2)(k_x+ik_y)^2$  in the off-diagonal
matrix elements of Eq. (\ref{eq:Luttinger}) are ignored. In this
approximation, $z-$projection of total angular momentum
${\mathcal{J}}_z=L_z+S_z$ is a conserved quantity.

We solve the Schr\"{o}dinger equation corresponding to Hamiltonian
Eq. (\ref{eq:Luttinger}), $H_L\Psi_{n,m,k_z}=E_{n,m,k_z}
\Psi_{n,m,k_z}$.  The eigenfunction is a 4 component spinor:
\begin{equation}
\label{eq:gen_wf} \Psi_{n, m, k_z}(r,\phi,z)\!=\!\sum_j\!\left(\!
\begin{array}{l}
a^{0,j}_{n,m,k_z} J_m(K_j r) e^{i m\phi}\\
a^{1,j}_{n,m,k_z} J_{m+1}(K_j r) e^{i (m+1)\phi}\\
a^{2,j}_{n,m,k_z} J_{m+2}(K_j r) e^{i (m+2)\phi}\\
a^{3,j}_{n,m,k_z} J_{m+3}(K_j r) e^{i (m+3) \phi}
\end{array}
\!\right)\!e^{ik_z \!z},
\end{equation}
where $m$ and $n>0$ are integers, $J_m$ is the Bessel function of
the first kind of order $m$. The radial wavevectors $K_j$ satisfy the
following secular equation:
\begin{widetext}
\begin{equation}
\label{eq:secular}\left|\!
\begin{array}{c c c c}
\frac{\gamma_1+\gamma_2}{2}K^2+\frac{\gamma_1-2\gamma_2}{2}k_z^2-E&
\sqrt{3}i \gamma_3 k_z K & \sqrt{3} \frac{\gamma_2+\gamma_3}{4} K^2 &0\\
-\sqrt{3}i \gamma_3 k_z K
&\frac{\gamma_1-\gamma_2}{2}K^2+\frac{\gamma_1+2\gamma_2}{2}k_z^2-E
&0&\sqrt{3} \frac{\gamma_2+\gamma_3}{4} K^2\\
\sqrt{3} \frac{\gamma_2+\gamma_3}{4} K^2&0&
\frac{\gamma_1-\gamma_2}{2}K^2+\frac{\gamma_1+2\gamma_2}{2}k_z^2-E&-\sqrt{3}i
\gamma_3 k_z K\\
0&\sqrt{3} \frac{\gamma_2+\gamma_3}{4} K^2&\sqrt{3}i \gamma_3 k_z
K&\frac{\gamma_1+\gamma_2}{2}K^2+\frac{\gamma_1-2\gamma_2}{2}k_z^2-E
\end{array}\!
\right|=0.
\end{equation}
\end{widetext}
The determinant has four positive roots $K_j>0$, $j=0,1,2,3$ and their opposites $-K_j$ are also the solutions. As
integer order Bessel functions $J_m(K r)$ and $J_m(-K r)=(-1)^m
J_m(K r)$ are not independent, only the positive $K$'s are
needed. Coefficients $a_i^j$ are determined from the boundary and
normalization conditions. The Dirichlet boundary conditions are written as 
\begin{equation}
\label{eq:gen_bound_cond} \sum_{j}a^{i,j}_{n,m,k_z} J_{m+i} (K_j
R)=0~,i=0,1,2,3.
\end{equation}
In order for the coefficients $a^{i,j}_{n,m,k_z}$ be non-zero, the
conditions
\begin{equation}
\label{eq:bc_Det} \det|J_{m+i} (K_j R)|=0~,
\end{equation}
have to be satisfied. Equations (\ref{eq:secular}) and
(\ref{eq:bc_Det}) uniquely determine the eigenenergies of the
problem. We note that the Kramers degeneracy occurs for states with indexes $m$
and $-3-m$.

\begin{figure}
\centering
\includegraphics[width=.5\columnwidth]{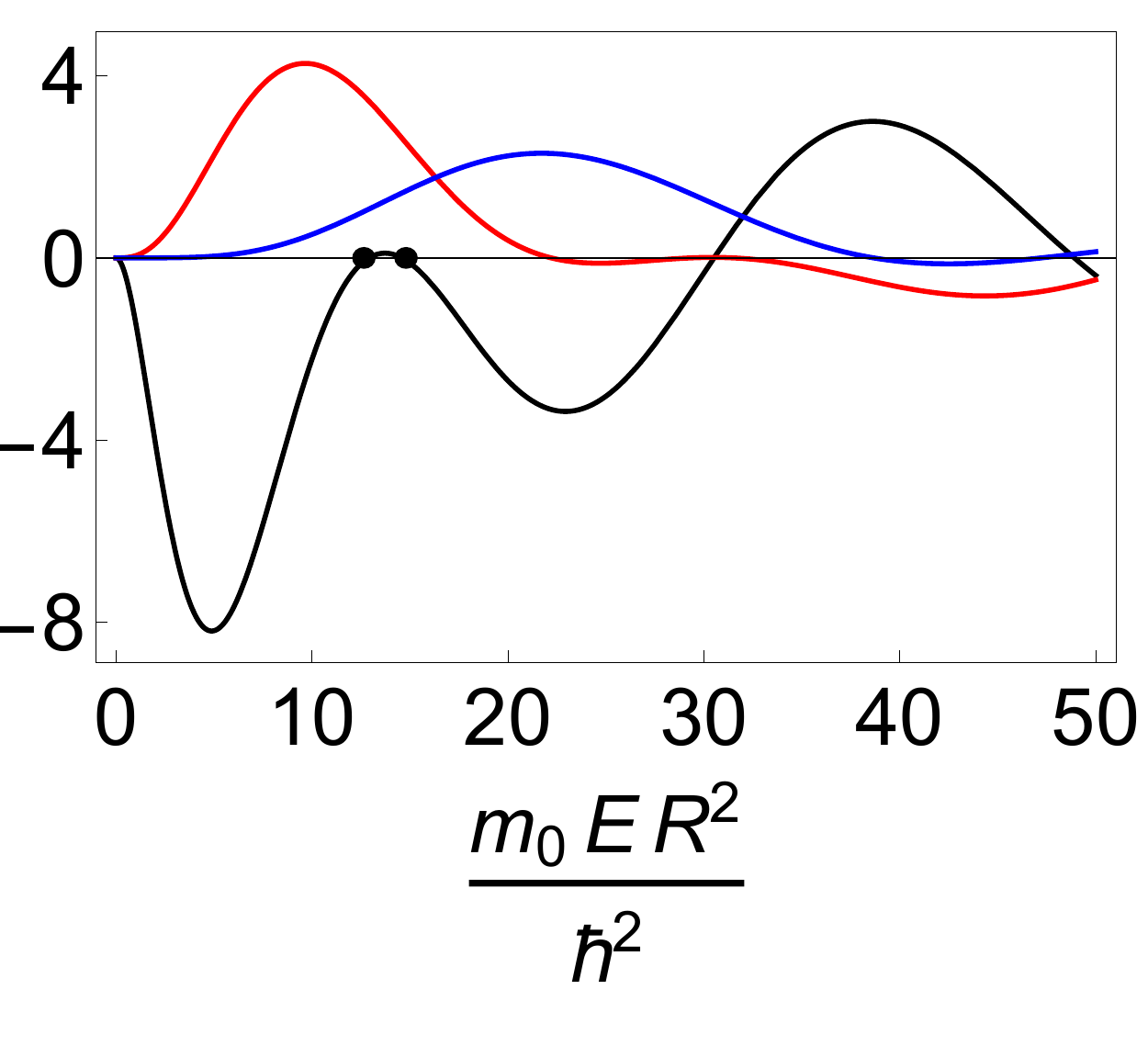}
\caption{.\label{fig:Bessel_sol} Plot of expression from Eq. (\ref{eq:zero_En}). Black-
$m=-2$, red - $m=0$, blue - $m=1$. Here $\gamma_1=6.8$,
$\gamma_2=2.1$, $\gamma_3=2.9$. Dots indicate the ground state and the first
excited state} 
\end{figure}

We solve the eigenvalue problem in the limit of small $k_z$. The
eigenevalues of $H_L$ can be expanded as:
$E_{n,m,k_z}=E_{n,m}^{(0)}+E_{n,m}^{(1)}k_z+ E_{n,m}^{(2)} k_z^2+...$. Using
this expansion, we compute $K_j$ from Eq. (\ref{eq:secular}),
retaining only terms up to $k_z^2$. We also expand  Eq. (\ref{eq:bc_Det})
in series of $k_z$ and solve the resulting equation for
$E_{n,m}^i$. Following this procedure, the zeroth order part of the
energy is obtained as a solution of the transcendental equation:
\begin{eqnarray}
\label{eq:zero_En} \left[ (\Gamma-2\gamma_2) \mathcal F_{m+1}^{m+3}+(\Gamma+2\gamma_2)\mathcal F_{m+3}^{m+1}\right]\times\nonumber\\
\left[ (\Gamma+2\gamma_2) \mathcal F_{m}^{m+2}+ (\Gamma-2\gamma_2)F_{m+2}^{m}\right]=0,
\end{eqnarray}
where
\begin{eqnarray}
\mathcal F_p^q&=&J_p\left[ {\mathcal
K}_2\left(E_{n,m}^{(0)} \right)R \right]J_q\left[{\mathcal
K}_1\left(E_{n,m}^{(0)} \right)R\right]\\
\Gamma&=&\sqrt{7\gamma_2^2+6\gamma_2\gamma_3+3\gamma_3^2}\\
{\mathcal K}_1(x)&=& \sqrt{\frac{4m_0 x}
{\hbar^2\left(2\gamma_1-\Gamma\right)}}\\
{\mathcal K}_2(x)&=& \sqrt{\frac{4m_0 x}
{\hbar^2\left(2\gamma_1+\Gamma\right)}}
\end{eqnarray}
A plot of the above expression and energies of the ground state and
the first excited state are shown in the Fig. \ref{fig:Bessel_sol}.
The ground state energy is obtained as the first solution of Eq.
(\ref{eq:zero_En}) for $m=-2$ and $m=-1$. We denote
$E_0=E_{0,-2}^{(0)}=E_{0,-1}^{(0)}$ for $m=-2$ and $m=-1$ . The second
solution of Eq. (\ref{eq:zero_En})is the first excited state.
The wavefunctions of the double-degenerate ground state are given by
\begin{eqnarray}
\label{eq:corr0_psi2} \Psi_{0,-2}^0&=&{\mathcal N}\left(
\begin{array}{c}
\frac{\Gamma-2\gamma_2}{\sqrt{3} \left(\gamma_2+\gamma_3\right)}
F_2(k_1,k_2,R,r) e^{-2i\phi}\\
0\\
F_0(k_1,k_2,R,r)\\
0
\end{array}
\!\right),\\
\label{eq:corr0_psi1} \Psi_{0,-1}^0&=&{\mathcal N}\left(
\begin{array}{c}
0\\
F_0(k_1,k_2,R,r)\\
0\\
\frac{\Gamma-2\gamma_2}{\sqrt{3} \left(\gamma_2+\gamma_3\right)}
F_2(k_1,k_2,R,r) e^{2i\phi}
\end{array}
\right),
\end{eqnarray}
where $k_i={\mathcal K}_i(E_0)$,  $\mathcal N$ is the
normalization factor and 
\begin{equation}
F_n(k_1,k_2,R,r)=J_n\left(k_1 r\right) -\frac{J_n\left(k_1 R\right)}
{J_n\left(k_2 R\right)}
J_n\left(k_2 r\right)
\end{equation}
 The dominant component in these wavefunctions
 corresponds to the angular momentum $\pm 1/2$.

The first order in $k_z$, the expansion  does not lead to any corrections to
energy, as it should be on symmetry grounds in the absence of the linear in hole momentum spin-orbit interactions; $E_{n,m}^{(1)}=0$. However corrections to the ground state wavefunctions
are non-zero and are given by
\begin{eqnarray}
\label{eq:corr1_psi2} \Psi_{0,-2}^1&=&+ iF_1(k_1,k_2,R,r) \left(
\begin{array}{c}
0\\
\delta_1 e^{-i\phi}\\
0\\
\delta_2 e^{i\phi}
\end{array}
\right)~,\\
\label{eq:corr1_psi1} \Psi_{0,-1}^1&=&-i F_1(k_1,k_2,R,r)\left(
\begin{array}{c}
\delta_2 e^{-i\phi}\\
0\\
\delta_1 e^{i\phi}\\
0
\end{array}
\right)~,
\end{eqnarray}
where
\begin{eqnarray}
\delta_1\!&\!=\!&\frac{\gamma_3}{k_1}
\frac{\Gamma-2\gamma_2}{\gamma_2+\gamma_3}
\frac{4\gamma_1+2\Gamma}{4\gamma_1\gamma_2-\Gamma^2} \left[1- 
\frac{k_2 J_1\left(k_1 R\right) J_2\left(k_2 R\right)}
{k_1 J_1\left(k_2 R\right) J_2\left(k_1 R\right)}\right]\nonumber \\ \\
\delta_2\!&\!=\!&\frac{4}{\sqrt{3}}\frac{\Gamma-2\gamma_2}{\left(\gamma_2+\gamma_3\right)^2
k_1}-
\frac{1}{\sqrt{3}}\frac{\Gamma-2\gamma_2}{\gamma_2+\gamma_3}\delta_1
\end{eqnarray}
Thus the two degenerate ground state wavefunctions are given by
\begin{eqnarray}
\label{eq:exp_wf_gs} \Psi_{0,-2,k_z}\approx\Psi_{0,-2}^0+k_z
\Psi_{0,-2}^1~,\\
\Psi_{0,-1,k_z}\approx\Psi_{0,-1}^0+k_z
\Psi_{0,-1}^1~.
\end{eqnarray}

The expansion of Eqs.
(\ref{eq:secular}) and (\ref{eq:bc_Det}) up to $k_z^2$ terms defines the effective mass of holes. A tedious but straightforward calculation gives the coefficient in front of $k^2_z$, $E_{n,m}^{(2)}$. The effective mass for motion
along the wire is $1/2 E_{n,m}^{(2)}$, $(\hbar=1)$. Its analytic expression is rather complicated but a simplified one can be written  for  $\gamma_2=\gamma_3$. Using the notations  
$\nu=(\gamma_1+2\gamma_2)/(\gamma_1-2\gamma_2)$ and $\tilde f_n^m =J_n(k_1) J_m(k_1 \sqrt{\nu})$, we obtain
\begin{eqnarray}
m=\left[\nu  k_1\tilde f_1^1 \left(-3 \tilde f_1^0  +\tilde f_1^2 -\sqrt{\nu }
\tilde f_0^1 +3 \sqrt{\nu } \tilde f_2^1 \right)\right]^{-1} \nonumber\\
\left[ 6 (\nu -1)   \tilde f_2^0  f_1^1 + \left(-3 \sqrt{\nu } \tilde f_1 ^0 +\sqrt{\nu }\tilde  f_1^2-\tilde f_0^1 +3 \tilde f_2^1\right)\right. \nonumber\\ 
\times\sqrt{\nu } k_1  \tilde f_1^1 -\left. 6 \left(\sqrt{\nu } \tilde f_1^0 \tilde f_1^2-(\nu +1) \tilde f_2^0 \tilde f_1^1
+\sqrt{\nu }\tilde f_0^1 \tilde f _2^1\right) \right]
\end{eqnarray}

\begin{figure}
\centering
\includegraphics[width=.72\columnwidth]{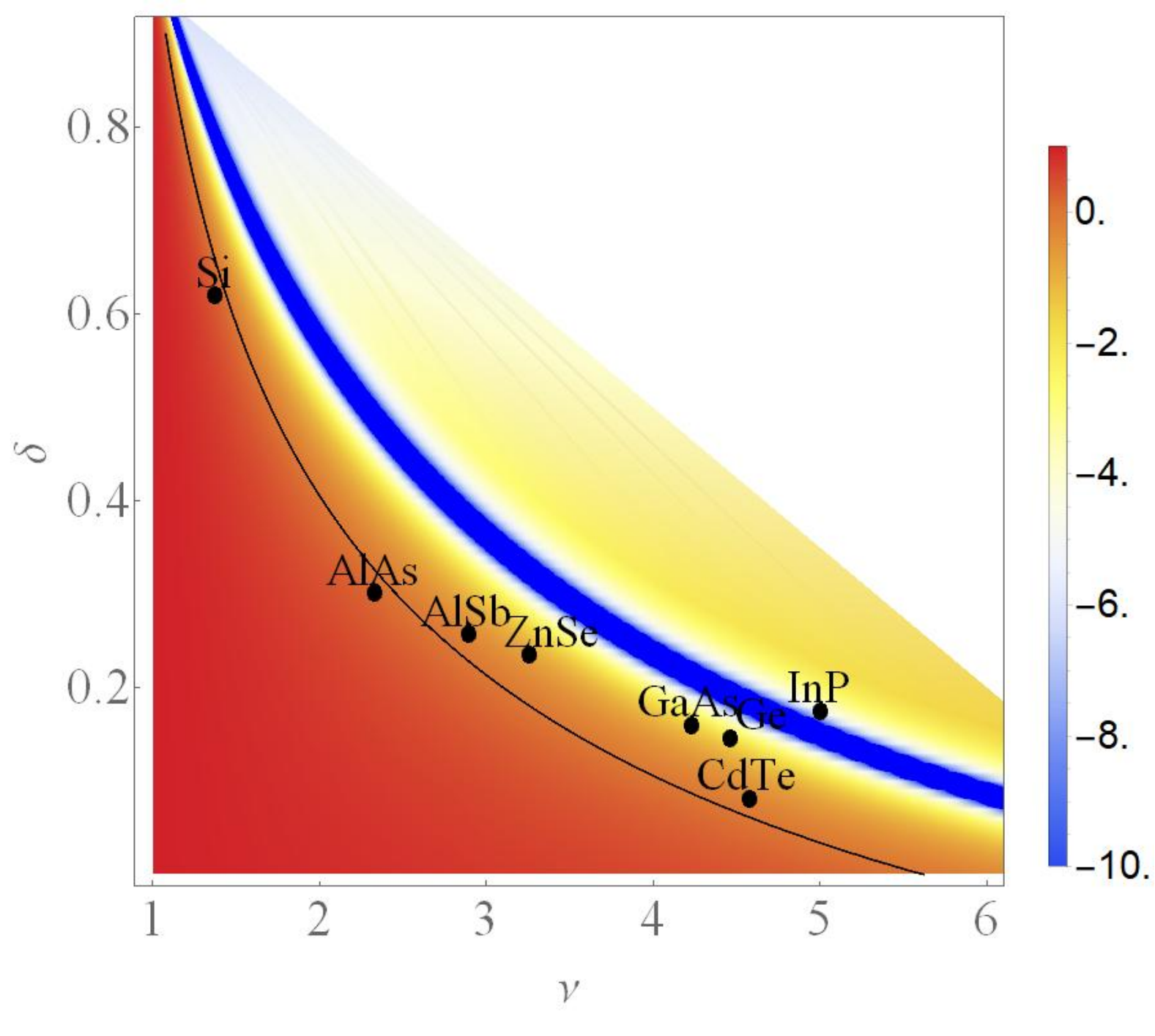}
\caption{Inverse effective mass as a function of the ratio of heavy and light bulk hole masses $\nu$ and anisotropy parameter $\delta$.
Black line corresponds to $1/m_{eff}=0$, while the dark blue region
indicates a very small mass, due to crossing between the ground state
and the first excited state. } \label{fig:mass_axial}
\end{figure}

The analysis of this expression shows that the effective mass can be infinite, so that there are no $k_z^2$
terms in the energy spectrum. In Fig. \ref {fig:mass_axial} we plot
the inverse effective mass as a function of the ratio of the bulk light and heavy hole masses $\nu$ and anisotropy
coefficient $\delta=(\gamma_3-\gamma_2)/(\gamma_3+\gamma_2)$. The
black line corresponds to the infinite mass. In the dark blue region the
mass is very small due to the crossing of the ground state and the first excited level. We have calculated the
effective masses for wires made of the several semiconductor compounds.

When the mass is infinite (or, in other terms, the inverse mass vanishes or is close to zero), the leading terms in the spectrum of holes are linear in
momentum and spin terms. These terms can emerge due to asymmetry of the confining potential of the wire or due to the bulk inversion asymmetry. We now consider asymmetry of the confining potential, which can be described by the presence of the electric field. The effect of an electric field $\boldsymbol{\mathcal E}$ that is perpendicular to the wire axis is described by Hamiltonian
\begin{equation}
H_e=e\mathcal Er \cos(\phi-\phi_0)~,
\end{equation}
where $\phi_0$ is the angle between $\boldsymbol{\mathcal E}$ and the $x$-axis.The  matrix
elements of $H_e$ on the two degenerate wavefunctions of the ground
state (expanded up to first order in $k_z$) are given by
\begin{eqnarray}
\braket{\psi_{0,-2,k_z}\left|H_e\right|\psi_{0,-2,k_z}}&=&\braket{\psi_{0,-1,k_z}\left|H_e\right|\psi_{0,-1,k_z}}=0\nonumber\\
\braket{\psi_{0,-2,k_z}\left|H_e\right|\psi_{0,-1,k_z}}&=&-i k_z
\beta \mathcal E \exp(i\phi_0)~,
\label{Rashba1}
\end{eqnarray}
where $\beta$ is evaluated using Eqs.
(\ref{eq:corr0_psi2}-\ref{eq:corr1_psi1}).
In addition to $H_e$, the electric field in the presence of mixing of higher bands in III-V or Ge and Si semiconductors leads to a standard spin-orbit Hamiltonian for bulk holes
\begin{equation}
H_{R}=\beta_R \mathcal E\left(J_x \cos \phi_0-J_y \sin \phi_0 \right)k_z
\label{bulk}
\end{equation}
where $\beta_R$ is the Rashba coefficient characterizing bulk interaction (\ref{bulk})  \cite{Jingcheng}. The matrix elements of this Hamiltonian between degenerate ground states of Eqs.
(\ref{eq:corr0_psi2}) and (\ref{eq:corr0_psi1}) are given by
\begin{eqnarray}
\braket{\psi_{0,-2,k_z}\left|H_R\right|\psi_{0,-2,k_z}}&=&\braket{\psi_{0,-1,k_z}\left|H_R\right|\psi_{0,-1,k_z}}=0\nonumber\\
\braket{\psi_{0,-2,k_z}\left|H_e\right|\psi_{0,-1,k_z}}&=&\beta_R {\mathcal E}{\mathcal {M}}e^{-i\phi_0} k_z
\label{Rashba2}
\end{eqnarray}
where
\begin{equation}
{\mathcal M}={\mathcal{N}}^2 \int_0^R {dr r } {\left[J_0\left(k_1
r\right) -\frac{J_0\left(k_1 R\right)} {J_0\left(k_2 R\right)}
J_0\left(k_2 r\right)\right]^2}.
\end{equation}

\begin{figure}
\centering
\includegraphics[width=.58\columnwidth]{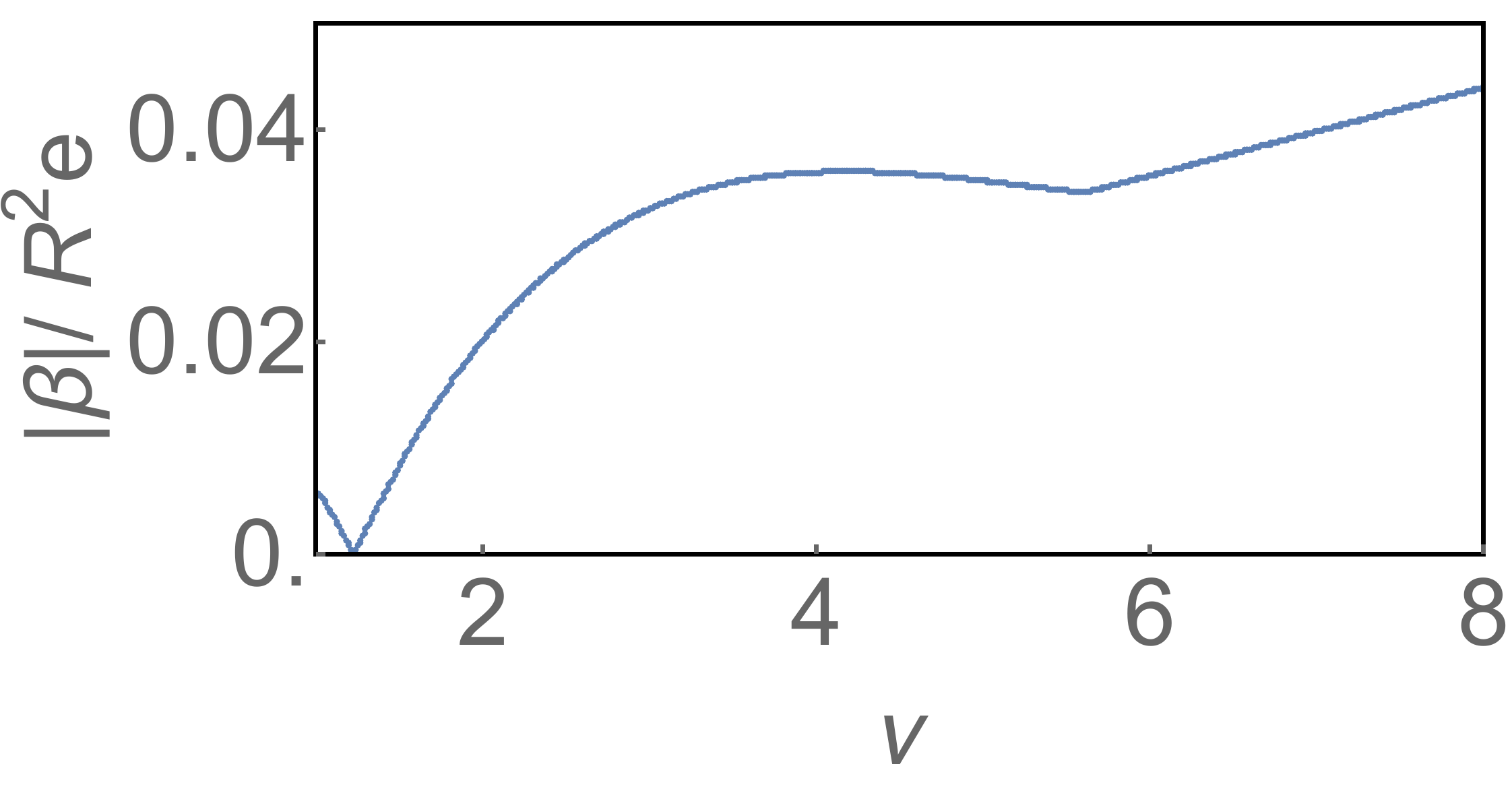}
\caption{Coefficient $\beta$ at the points of infnite mass} \label{fig:beta_coeff}
\end{figure}

In the Hilbert space of the two degenerate
ground state wavefunctions $\psi_{0,-1,k_z}$ and $\psi_{0,-2,k_z}$,
$H_e +H_R$ acts as an effective Rashba and Zeeman Hamiltonian
\begin{equation}
H_{eff}^{RZ}=\beta k_z \boldsymbol{\mathcal E} \times \boldsymbol \sigma +\beta_R \mathcal M k_z  \boldsymbol{\mathcal E}\cdot\boldsymbol \sigma~.
\label{eff}
\end{equation}
When the inverse mass vanishes, matrix element $\mathcal M$ is very close to unity while coefficient $\beta$ depends on the radius of the wire and is plotted in Fig. \ref{fig:beta_coeff}

The wire Hamiltonian is similar to Hamiltonian describing an edge in topological insulators \cite{fu2008superconducting,alicea2012new},
\begin{equation}
{\cal H}_{w} =\int dz\Psi^{\dagger}\left(-iv\partial_z -\mu\right) \Psi,
\end{equation}
where velocity $v$ is defined as
\begin{equation}
\hbar v=\mathcal E \sqrt{\beta_R^2 \mathcal M^2 +\beta ^2 +2 \beta 
\beta_R \mathcal M \cos \left(2\phi_0\right)}
\end{equation}
 and $\Psi^{\dagger}_z$ adds an electron with spin $\sigma$ at a coordinate $z$.
In the presence of superconducting proximity effect described by the induced order parameter $\Delta$, which couples charge carriers with opposite spins, given by
\begin{equation}
{\cal H}_{\Delta}= \int dz \Delta \left( \Psi_{\downarrow}\Psi_{\uparrow}+H.C\right),
\end{equation}
 the Bogoliubov-DeGennes Hamiltonian ${\cal H}_{BdG}={\cal H}_{w}+{\cal H}_{\Delta}$ leads to quasiparticle energies
\begin{equation}
E_{\pm}(k)= \sqrt{(\pm vk-\mu)^2)+\Delta^2},
\end{equation}
describing a gapped topological superconductor. In contrast to conventional semiconducting wires with shifted parabolic Rashba/Dresselhaus spectrum, where topological superconductivity and Majorana fermions at the end of the wire arise at Zeeman splitting of order to superconducting gap, here they emerge even at zero magnetic field. In a finite magnetic field $\mathbf{H}$ ,
e.g., along $z-axis$, the quasiparticle energies are given by
\begin{equation}
\label{energies}
E_{\pm}(k)= \sqrt{\frac{\epsilon_+^2+\epsilon_-^2}{2}\pm\left(\epsilon_+-\epsilon_-\right)\sqrt{\Delta_s^2+\mu^2}+\Delta^2},
\end{equation}
where $\epsilon_{\pm}=-\mu\pm \sqrt{(vk)^2 +Z^2 }$, $Z=g\mu_BH$ is the Zeeman energy, $\mu_B$ is the Bohr magneton and $g$ is the longitudinal $g-$factor along the direction of the wire, $\Delta_s=Z\Delta/ \sqrt{(vk)^2 +Z^2} $ is the strength of induced s-paring in the wire. Notably, $\Delta_s$ is nonzero only in the presence of magnetic field. The quasiparticle gap
vanishes at magnetic field corresponding to $Z=\Delta^2 + \mu^2$. At a higher magnetic fields the gap reopens, but the the superconductivity is no longer topological, while at a smaller fields the superconductivity is topological. If the proximity pairing amplitude $\Delta$ varies along the wire, then at a certain magnetic field, the topological superconductor will be in the region of the wire with $\Delta > Z$, while the conventional superconductor will be in the region with $\Delta < Z$. Tuning $Z$ can make possible moving the Majorana zero modes, localized at the boundaries between topological and non-topological regions. In the Rashba wires with parabolic spectrum, the situation is the opposite: the topological superconductivity persists at the  Zeeman energy larger than $\Delta^2 + \mu^2$.

It is known that parameters of the valence band, such as masses can be tuned not only varying the composition of heterostructure materials, but also tuning external parameters, e.g. such as strain. This potentially opens a possibility to induce an infinite mass in the wire by strain, and to observe transition between the cases of Rashba wire and spin-locked wire.

\section{Acknowledgement}
 This work is supported by the U.S. Department of Energy, Office of Basic Energy Sciences, Division of Materials Sciences and Engineering under Award DE-SC0010544.

%\bibliography{bibtit2}
\end{document}